\documentclass[twocolumn,showpacs,pra]{revtex4}
\usepackage{graphicx}
\usepackage{amssymb}

\usepackage{dcolumn}
\usepackage{bm}
\usepackage{amsmath}
\begin{document}
\title{Effects of Zero Mode and Thin Spectrum on the Life Time of Atomic Bose Einstein Condensates}
\author{T. Birol}
\affiliation{Department of Physics, Cornell University, Ithaca,
14853 NY, U.S.A.}
\affiliation{Department of Physics, Ko\c{c}
University, Sar{\i}yer, 34450, \.{I}stanbul, Turkey}
\author{\"O. E. M\"ustecapl{\i}o\u{g}lu}
\affiliation{Department of Physics, Ko\c{c} University, Sar{\i}yer,
Istanbul, 34450, Turkey}
\begin{abstract}
Reviewing the ideas developed in \cite{myself}, the ground state
life time of a finite size atomic Bose Einstein condensate is
studied for coherent, squeezed coherent and thermal coherent ground
states. Ground state evolution of coherent and squeezed coherent
states in a double well potential is studied. Effects of thin
spectrum on Bose-Einstein condensates is discussed and quasiparticle
excitation lifetimes are calculated. It is shown that the effect of
the states we use on the free energy vanishes in the thermodynamic
limit. Possible extension to a double well potential and effect of a
second broken symmetry is also discussed. \end{abstract}

\maketitle
\section{Introduction}
Shortly after Bose Einstein condensation \cite{bec_original} was
obtained in trapped Alkali atoms \cite{bec95}, many theoretical and
experimental studies focused on the quantum coherence properties of
such systems. It was shown that apart from the usual decoherence,
which stems from the imperfect isolation from the environment, the
system also suffers phase diffusion \cite{walls,you96}, which is due
to the atomic number fluctuations in the condensate \cite{hansch}.
There is a third source of decoherence which limits the life time of
excitations in BECs. This mechanism is based on the existence of a
group of \textit{thin spectrum} states \cite{wezel06}. The relation
of thin spectra with decoherence of excitations is recently proposed
and applied to Lieb-Mattis model and superconductors by the same
authors \cite{wezel05,wezel07}. Effect of a thin spectrum on
quasiparticle excitations on BECs is discussed in \cite{myself}. In
this paper we review these ideas with some extensions and also
discuss the decoherence that a double well BEC experiences.

This paper is organized as follows: Section II begins with a review
of a toy model for zero mode dynamics, studying coherent, squeezed
coherent and thermal coherent ground state lifetimes. A toy model
for a double well condensate is also introduced and coherent and
squeezed coherent states in such a system is studied. In section III
we apply the thin spectrum formalism to atomic BECs and discuss its
similarities and differences with other calculations. After making
some comments about the existence of a second thin spectrum in a
double well system and outlining the calculation of lifetime in a
double well system, we conclude in section IV. Acknowledgements are
in section V.

\section{Toy Model}
\subsection{Introduction}
 In order to understand the basic idea underlying the phase
diffusion at zero temperature, it is useful to introduce a toy model
\cite{toy1,toy2}. The total Hamiltonian of a homogeneous Bose
Einstein condensate in the weakly interacting limit is
\begin{equation}
\mathcal{H}= \sum_k E_k a^\dagger_k a_k + \frac{\tilde u}{2}\sum_{k,
p, q}a_{p+q}^\dagger a_{k - q}^\dagger a_k a_p \label{equ_ham1}
\end{equation}
where $a_k$ is the annihilation operator of the $k$ mode and
$\tilde{u}$ is the parameter determining the strength of
interactions between bosons, that is $\tilde{u}=\frac{4 \pi
\hbar^2 a_s}{m V}$, where $V$ is the quantization volume. In order
to fix the average number of atoms, a chemical potential $\mu$ is
also included in single particle energy: $E_k=\frac{\hbar^2 k^2}{2
m} - \mu$.

We are interested in the zero-mode dynamics of the system, so we
discard terms including $a_{\vec{k}\neq 0}$ and redefine $\mu$ in
order to get the basic $U(1)$ gauge symmetric Hamiltonian:
\begin{eqnarray}
\mathcal{H}=\frac{\tilde{u}}{2}a^\dagger a^\dagger a a - \mu
a^\dagger a, \label{u1hamiltonian}
\end{eqnarray}
The grounstate of such a Hamiltonian is clearly a Fock state with a
number $N$ determined by $\mu$. However, Fock states have no
definite phase and since Bose Einstein condensation entails a
phase-symmetry broken state, the groundstate we seek cannot simply
be a Fock state. The simplest idea is to consider a coherent state
$|\alpha\rangle$ with $\alpha= \sqrt{N}$, as coherent states are the
simplest states carrying a (almost) definite phase. A second step
can be considering squeezed coherent states \cite{mandel}, which
again carry some phase information, but the uncertainty in their
phase can be larger or less than a corresponding coherent state. And
finally, it is important to find a state which both carries phase
information and is temperature dependent. For this purpose, we are
going to introduce the \textit{thermal coherent states} after
studying the zero-mode evolutions of coherent and squeezed coherent
states.

\subsection{Lifetime for coherent groundstate}
The coherent (or quasi-classical \cite{cohentannoudji}) state $|
\alpha \rangle$ is defined as the right eigenstate of the
annihilation operator and has the Fock state expansion
\begin{equation}
|\alpha\rangle = {\rm e}^{-|\alpha|^2 /2} \sum_{n=0}^{\infty}
\frac{\alpha^n}{\sqrt{n!}} |n\rangle .
\end{equation}
This simple expansion makes it possible to calculate the time
dependence of the expectation value of the annihilation operator
$a$, which we consider as the order parameter, possible. We define
the energy of the $n$-{th} Fock state $|n\rangle$ as
$E_n=\frac{\tilde{u}}{2}(n^2 - n) - \mu n$ through
$\mathcal{H}|n\rangle = E_n |n\rangle$. This gives the simple time
dependent expression
\begin{eqnarray}
\langle \alpha|a|\alpha\rangle
    &=&\sqrt{N} \exp\left(N[{\rm e}^{-\frac{i}{\hbar}
    \tilde{u}t}-1]\right){\rm e}^{\frac{i}{\hbar}\mu t},
\label{gte}
\end{eqnarray}
whose short time behavior is found to be
\begin{eqnarray}
\langle \alpha|a|\alpha\rangle = \sqrt{N} {\rm
e}^{\frac{i}{\hbar}\mu t}{\rm e}^{-i\frac{N\tilde{u}}{\hbar}t} {\rm
e}^{-\frac{N \tilde{u}^2}{2\hbar^2} t^2}.
\end{eqnarray}
Therefore, the order parameter decays exponentially
\cite{walls,you96}. At longer time scale $\langle
\alpha|a|\alpha\rangle$ revives due to discrete and thus periodic
nature of the exact time evolution (\ref{gte}). However, the ratio
of the revival time $t_r$ to the collapse time scales as $t_r/t_c
=\sqrt{N}$, and hence in the thermodynamic limit the collapse is
irreversible. Denoting the density of condensed atom numbers in the
quantization volume as $\rho=N/V$, collapse time
$t_c=\hbar/\sqrt{N}\tilde{u}$ can be written as
$t_c=\hbar\sqrt{N}/\rho u_0$ to see its behavior in the
thermodynamic limit more directly. As $\rho$ is fixed in the
thermodynamic limit where $N$ and $V$ increases indefinitely, we see
that $t_c$ increases indefinitely. Revival time
$t_r=\hbar/\tilde{u}=\hbar N/u_0\rho$ increases with $N$ linearly.
In practice, the available condensates contain small number of atoms
and furthermore, they are in traps that makes them inhomogeneous
systems. Our homogeneous system Hamiltonian can qualitatively
describe their collapse time behavior by letting $V$ denote the
condensate mode volume, though $t_c,t_r$ would have different
expressions for the case of a trapped condensate. In particular,
collapse time of a harmonically trapped three dimensional isotropic
condensate in a coherent state behaves like $t_c\sim N^{1/10}$ in
the Thomas-Fermi limit \cite{toy2}. Homogenous condensate collapse
time is growing much faster, as $t_c\sim N^{1/2}$. Despite these
quantitative differences, we can still express $t_c$ of homogeneous
BEC in terms of parameters of a trapped BEC. For that aim we shall
only eliminate $m$ via the characteristic length scale for a
harmonic trap potential as $a_{\rm ho}=\sqrt{\hbar/m \omega_{\rm
tr}}$ in terms of the harmonic trap frequency $\omega_{\rm tr}$. We
find
\begin{eqnarray}
t_c=\frac{\sqrt{N}}{4 \pi N_{\rm eff}}\frac{1}{\omega_{\rm tr}},
\end{eqnarray}
where $N_{\rm eff}=\rho a_{\rm ho}^2 a_s$. Assuming a typical
situation of current experiments with $N\sim 10^6$, $a_s=10$ nm,
$a_{\rm ho}=1$ $\mu$m, and $\rho=10^{21}$ m$^{-3}$, we get
$t_c\simeq 10/\omega_{\rm tr}$. For a harmonic trap with
$\omega_{\rm tr}=100$ Hz, this amounts to $t_c\sim 10^{-1}$
seconds, clearly within the regime to be confirmed and studied
experimentally \cite{hansch}.

\subsection{Lifetime for squeezed coherent groundstate}
The squeezed coherent state \cite{mandel} $|\alpha, \gamma\rangle$
is defined as $|\alpha, \gamma\rangle =
D(\alpha)S(\gamma)|vac\rangle$ where
\begin{eqnarray}
S(\gamma)={\rm e}^{\frac{\gamma}{2} a a- \frac{\gamma^*}{2}
a^\dagger a^\dagger}.
\end{eqnarray}
is the unitary squeezing operator and $D(\alpha)=\exp(\alpha a
-\alpha^* a^\dagger)$ is the displacement operator. This again is a
minimum uncertainty state, but the quantum fluctuations of two
quadratures are not equal to each other. Arguments of $\gamma$ and
$\alpha$ determine which quadrature is \textit{squeezed} at the
expense of increased uncertainty of the other one. In particular, if
both parameters are real and positive, then the state is
\textit{number squeezed}, that is the uncertainty of the number
operator is reduced whereas the conjugate variable, phase, has a
higher uncertainty. Such a state resembles a Fock state more than a
coherent state and therefore is expected to have a longer life time,
since the phase collapse speed is generally proportional to $\Delta
N$, which is smaller in this case, as have recently observed
experimentally \cite{mara,ibloch}. This situation is analogous to
the dispersion of a wavepacket consisting of different frequency
components.

\begin{figure}[h]
\centering{\vspace{0.5cm}}
\includegraphics[width=3.5in]{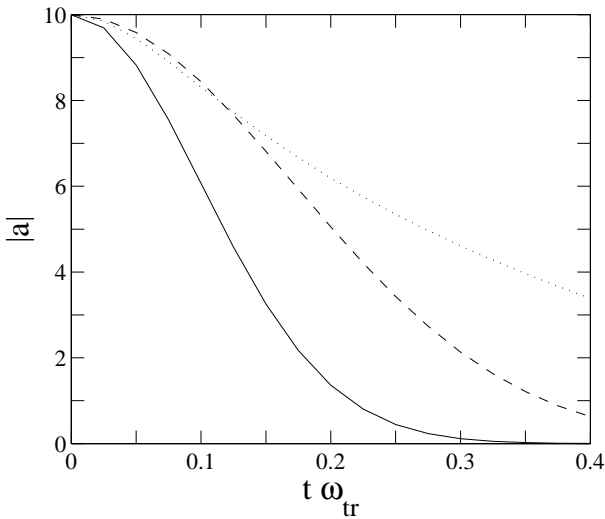}
\caption{The comparison of the short time decay character for a
coherent state condensate with that for squeezed states with
$\zeta=0.5$ and $\zeta=0.9$. Sample parameters used are $a_s=10nm$,
$a_{\rm ho}=1$ $\mu$m, $n=10^{21}$ m$^{-3}$, and $\alpha=10$
corresponds to $N=100$. In this case, the dimensionless time is in
units of $\hbar/\tilde{u}$ becomes $\hbar/\tilde{u}= \omega_{\rm
tr}^{-1}$. The fastest decay (solid line) is for the coherent state,
while the dashed (dotted) line refers to that of a squeezed state
with $\zeta=0.5$ ($\zeta=0.9$). \label{figstm1}}
\end{figure}

\begin{figure}[h]
\centering{\vspace{0.5cm}}
\includegraphics[width=3.5in]{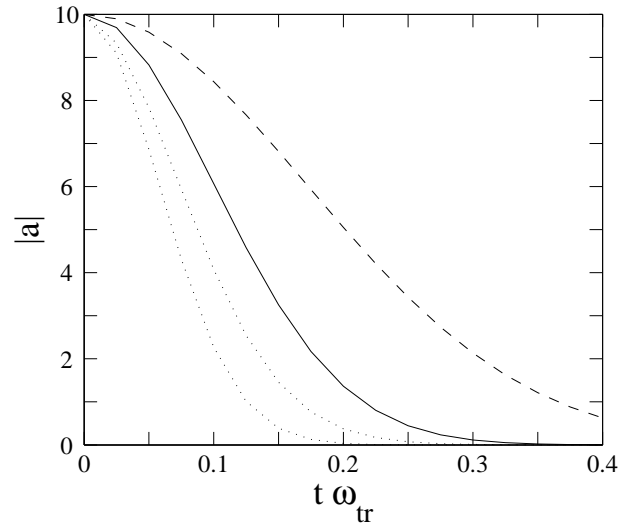}
\caption{Decay of the order parameter for the coherent state and
squeezed states of $\zeta=0.5$, $\zeta=0.5 i$ and $\zeta=-0.5$ as a
function of $t \omega_{tr}$. The same sample parameters with the
previous plot are used. The solid line is the coherent state, the
dashed line is the squeezed state with $\zeta=0.5$, and the dotted
ones are the squeezed states with $\zeta=0.5i$ and $\zeta=-0.5$.
\label{figstm2}}
\end{figure}

In terms of a new parameter
\begin{eqnarray}
\zeta= \gamma \frac{\tanh(|\gamma|)}{|\gamma|}
\end{eqnarray}
the Fock state expansion of the squeezed coherent state is
\cite{wunsche}
\begin{eqnarray}
|\alpha, \gamma \rangle
    &=& \sum_{n=0}^{\infty} A_n(\alpha, \zeta) |n\rangle\nonumber\\
    &=& (1-|\zeta|^2)^{1/4}
    {\rm e}^{- \frac{(\alpha + \zeta \alpha^*)\alpha^*}{2}}
    \sum_{n=0}^{\infty}
    \sqrt{\frac{\zeta^n}{2^n n!}} H_n\left(\frac{\alpha+\zeta \alpha^*}{\sqrt{2 \zeta}}\right)
    |n\rangle.
\end{eqnarray}
Here, $H_n$ is the n-th Hermite Polynomial. The order parameter
becomes
\begin{eqnarray}
\langle \alpha, \gamma | \hat{a}(t) | \alpha,\gamma \rangle =
\sum_{n=0}^{\infty}\sqrt{n+1}A_n^* A_{n+1}
 {\rm e}^{\frac{i}{\hbar}(E_n - E_{n+1})t},
 \label{esg}
\end{eqnarray}
which is not possible to evaluate analytically. We therefore attack
the problem using simple numerical methods, and plot the time
dependence of the order parameter for various values of $\zeta$,
both real and imaginary. It is clearly seen in figure \ref{figstm1}
that squeezing in the number direction increases the life time,
whereas in figure \ref{figstm2} we see squeezing in phase direction
leads to a faster decay.

\begin{figure}[h]
\centering{\vspace{0.5cm}}
\includegraphics[width=3.5in]{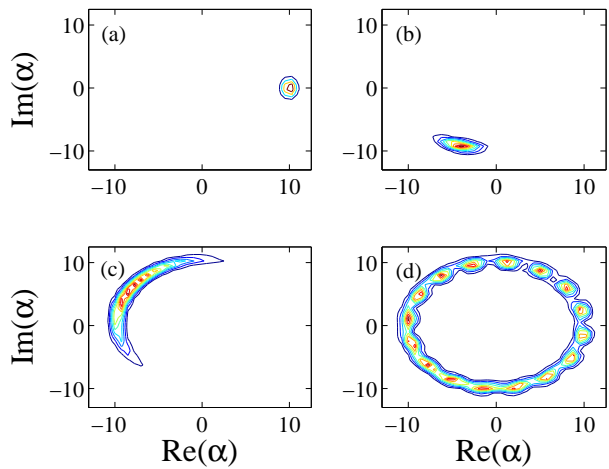}
\caption{Time evolution of the Q-function for squeezed-coherent
state with $\alpha=10$ and $\zeta=0.5$ for increasing values of $t
\omega_{tr}$. Figures (a), (b), (c) and (d) correspond to $t
\omega_{tr}=0$, $t \omega_{tr}=0.02$, $t \omega_{tr}=0.10$ and $t
\omega_{tr}=0.40$.
 \label{scsQ1}}
\end{figure}

\begin{figure}[h]
\centering{\vspace{0.5cm}}
\includegraphics[width=3.5in]{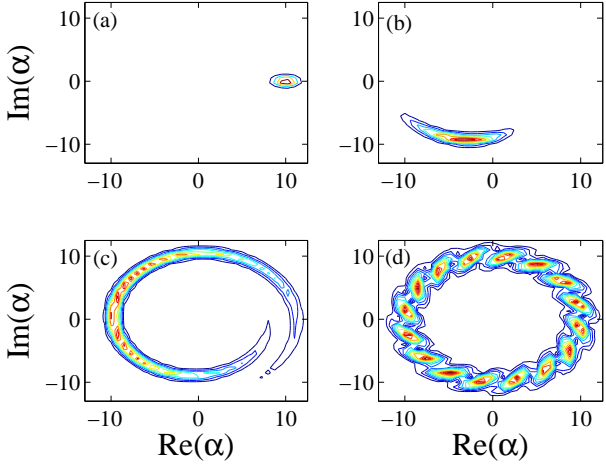}
\caption{Time evolution of the Q-function for squeezed-coherent
state with $\alpha=10$ and $\zeta=-0.5$ for increasing values of $t
\omega_{tr}$. Figures (a), (b), (c) and (d) again corresponds to
same time steps; $t \omega_{tr}=0$, $t \omega_{tr}=0.02$, $t
\omega_{tr}=0.10$ and $t \omega_{tr}=0.40$. The speed of phase
diffusion is higher\ since the state is a phase-squeezed state.
 \label{scsQ2}}
\end{figure}

Considering the time evolution of the Q-functions might provide some
extra insight into the phase diffusion. In a contour plot of a Q
function, radial distribution corresponds to the number distribution
whereas angular one gives the phase information. Beginning with a
state with some phase information, we expect it to get a
rotationally symmetric form as time passes and phase diffusion
occurs. This is seen in the figures \ref{scsQ1} and \ref{scsQ2}
which correspond to squeezed coherent states with $\alpha=10$ and
$\zeta=\mp0.5$ respectively.


\subsection{Lifetime for thermal coherent groundstate}
\begin{figure}[h]
\centering{\vspace{0.5cm}}
\includegraphics[width=3.5in]{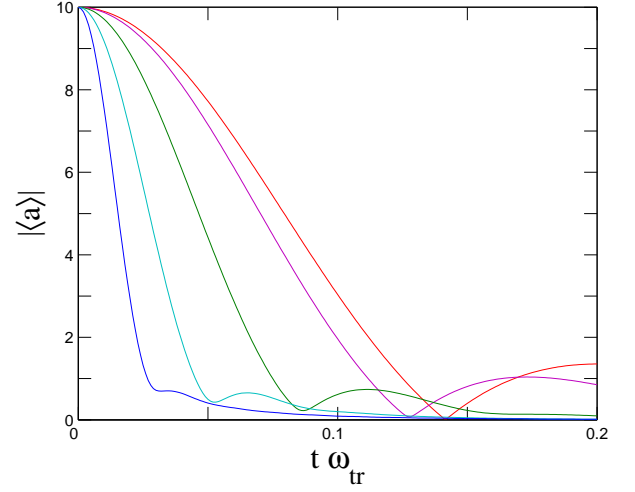}
\caption{The short time decays for thermal coherent states. The
lines correspond respectively to $T=1000$ nK, 100 nK, 10 nK, 1 nK,
and 0.001 nK from left to right. The humps are due to the ground
degeneracy $E_0=E_1$. Even as the temperature approaches zero, the
state does not approach the ordinary coherent state $D(\alpha)
|0\rangle$. Instead, it approaches a superposition state $D(\alpha)
(|0\rangle + |1\rangle)/\sqrt{2}$. It can be seen that the envelope
of the function for small $T$ decays at the same time scale as a
coherent state.\label{dtsop}}
\end{figure}

In order to study groundstate lifetime in finite temperature, we
need a state with both a thermal characteristic and phase
information. The thermal state with the density matrix
\begin{eqnarray}
\rho_{\rm th}   &=& {\rm e}^{-\beta \mathcal{H}}\nonumber\\
            &=& \sum_n {\rm e}^{- \beta E_n} |n\rangle\langle n|,
\label{eqt}
\end{eqnarray}
has a uniform phase, that is $\langle a \rangle = 0$. We introduce
the thermal coherent state as   $\rho = D(\alpha) \rho_{\rm th}
D^\dagger (\alpha)$, where $\rho_{\rm th}$ is defined using a
Hamiltonian without free energy, that is
$E_n|n\rangle=(\tilde{u}/2) a^\dagger a^\dagger a a |n\rangle$.
Using the Fock state expansion of displaced number state
$|n,\alpha\rangle$ \cite{roy82}
\begin{eqnarray}
    D(\alpha)|n\rangle  &=& |n, \alpha\rangle \nonumber\\
                        &=& \sum_{m=0}^{\infty} {\rm e}^{- \frac{1}{2}
                        |\alpha|^2} \sqrt{\frac{n!}{m!}}
                        \alpha^{m-n} L_n^{m-n}(|\alpha|^2)|m\rangle \hskip 24pt\nonumber\\
                        &=& \sum_{m=0}^{\infty} C_m (n, \alpha)
                        |m\rangle,
\end{eqnarray}
where $L_k^l$ are the generalized Laguerre polynomials, $\rho$
becomes
\begin{eqnarray}
    \rho    &=&\sum_{n m m'} {\rm e}^{-\beta E_n} C_m(n,\alpha)
            C_{m'}^*(n,\alpha) |m\rangle \langle m'|.
\end{eqnarray}
The order parameter, which is found by taking the trace of $a \rho$\
equals
\begin{eqnarray}
\langle a(t) \rangle    &=& \sum_{nmm'k} {\rm e}^{-\beta E_n}
C_m(n,\alpha) C_{m'}^\dagger(n,\alpha) \langle k|m\rangle \langle
m'|{\rm e}^{\frac{i}{\hbar}\mathcal{H}t}a {\rm
e}^{-\frac{i}{\hbar}\mathcal{H}t}|k\rangle ,\\
    &=& \sum_{n m} {\rm e}^{-\beta E_n}
C_{m+1}(n, \alpha) C_{m}^*(n, \alpha) \sqrt{m}\, {\rm
e}^{-\frac{i}{\hbar}(E_{m+1}-E_m) t}.
\end{eqnarray}
There will be destructive interference since $E_{n+1} - E_n$ is not
constant. The number of contributing factors is determined by
$\beta$, that is, temperature. Increased temperature will make more
terms contribute and hence lead to shorter lifetime. Time evolution
of the order parameter is plotted in figure \ref{dtsop} for
$\alpha=10$.

\begin{figure}[h]
\centering{\vspace{0.5cm}}
\includegraphics[width=3.5in]{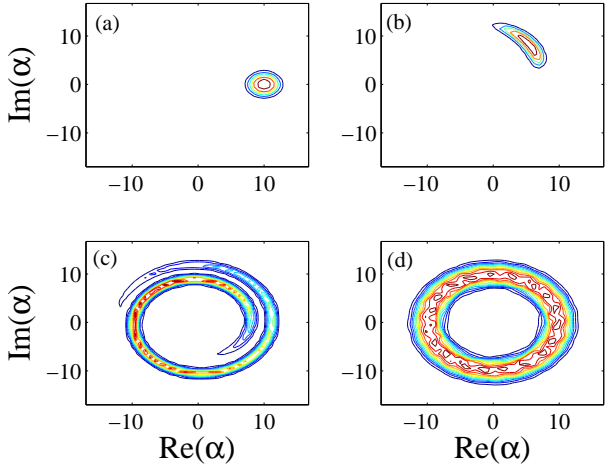}
\caption{Time evolution of the Q-function of a thermal coherent
state with $\alpha=10$ at $T=10$ nK. Figures (a), (b), (c) and (d)
correspond to $t \omega_{tr}=0$, $t \omega_{tr}=10^{-4}$,
$t=\omega_{tr}=10^{-3}$ and $t \omega_{tr}=10^{-2}$.}
\end{figure}
\begin{figure}[h]
\centering{\vspace{0.5cm}}
\includegraphics[width=3.5in]{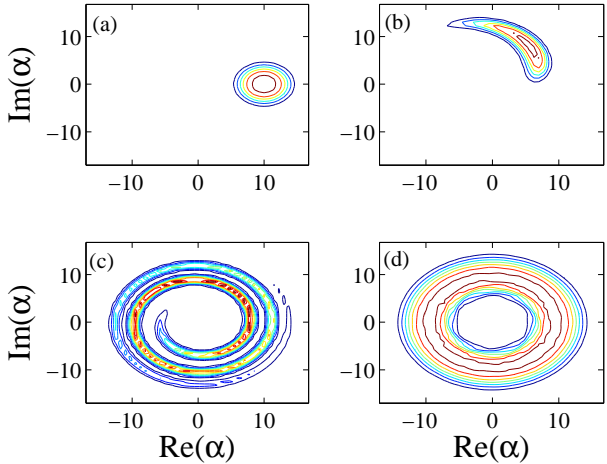}
\caption{Time evolution of the Q-function of a thermal coherent
state with $\alpha=10$ at $T=100$ nK. Figures (a), (b), (c) and (d)
correspond to $t \omega_{tr}=0$, $t \omega_{tr}=10^{-4}$, $t
\omega_{tr}=10^{-3}$ and $t \omega_{tr}=10^{-2}$. Wider radial
distribution shows that even when the phase information lost, there
is a higher uncertainty in number due to higher temperature.}
\end{figure}

\subsection{Toy Model for Double Well Potential}
We now consider the case when two condensates in identical potential
wells are brought into contact via a Josephson-like junction. The
toy model Hamiltonian is of the form \cite{toy1,toy2}
\begin{equation}
\mathcal{H}=\frac{\tilde{u}}{2}(a^\dag a^\dag a a + b^\dag b^\dag b
b)-\mu(a^\dag a + b^\dag b) - \lambda (a^\dag b + b^\dag a)
\end{equation}
where $a$ ($b$) is the annihilation operator for the zero mode of
the condensate in well A (B). We denote the state which has $n$
atoms in well A and $m$ atoms in well B by $|n, m\rangle$. Assuming
the total number of atoms in both wells is fixed and equal to $N$,
the ground state $|gr\rangle$ of the double well condensate can be
expanded as
\begin{equation}
|gr\rangle = \sum_{n=0}^\infty c_n |n, N-n\rangle .
\end{equation}
For identical wells, $c_n$ is expected to be peaked around $n=N/2$.
The strength $\lambda$ of the coupling determines the dispersion of
the number of atoms in a well. In particular, if one makes the
ansatz
\begin{equation}
|c_n|^2 \propto {\rm e}^{- \frac{(n-N/2)^2}{2 \sigma^2 (N/2)}}
\end{equation}
expanding the Schroedinger equation gives \cite{toy1,toy2}
\begin{equation}
\sigma^2=\frac{N}{4}\sqrt{\frac{\lambda}{N \tilde{u}/2 +\lambda}}.
\end{equation}

In order to study the phase collapse, we consider the correlation
$G=\langle gr|b^\dag a |gr\rangle$. The time evolution of this
expectation value after the coupling is turned off can be easily
shown to be
\begin{equation}
G=\sum_{n=0}^\infty {\rm e}^{-\frac{i}{\hbar} (E_n -E_{n+1}) t}
c_n^* c_{n+1} \sqrt{(n+1)(N-n)}.
\end{equation}
We have defined $\mathcal{H}|n,N-n\rangle =E_n|n,N-n\rangle$, so
$E_n=\frac{\tilde{u}}{2} \left(n (n-1) + (N-n)(N-n-1)\right)-\mu N$.
Plotting this for coherent, number squeezed and phase squeezed
states we see that the qualitative results we got from single well
toy model is still valid, as expected.

\begin{figure}[h]
\centering{\vspace{0.5cm}}
\includegraphics[width=3.5in]{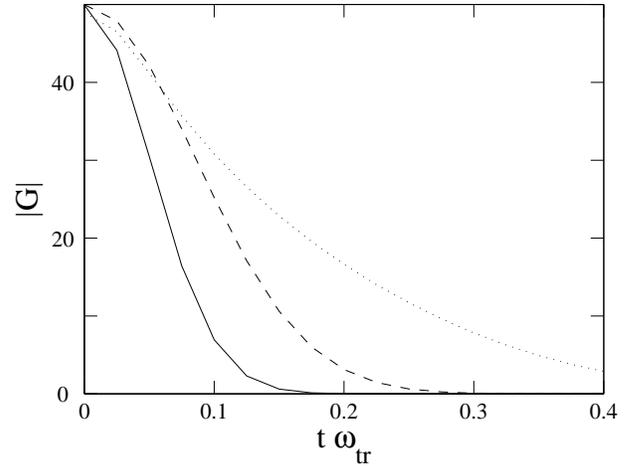}
\caption{Comparison of the short time decay for a double well
condensate which has 200 atoms equally distributed to two wells.
Solid line corresponds to a coherent state, dashed line corresponds
to a squeezed state with $\zeta=0.5$ and dotted line to a squeezed
state with$\zeta=0.9$.}
\end{figure}
\begin{figure}[h]
\centering{\vspace{0.5cm}}
\includegraphics[width=3.5in]{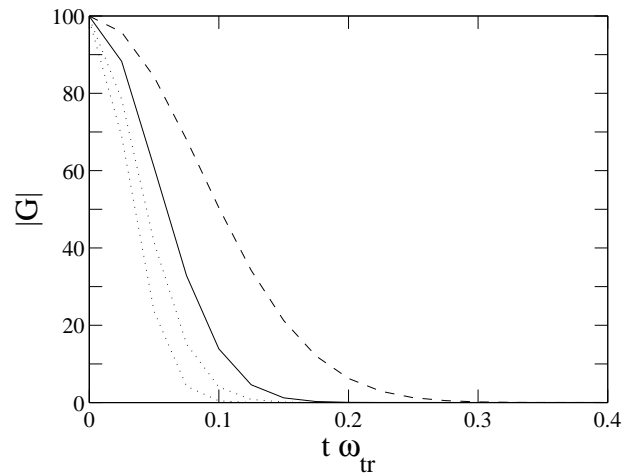}
\caption{Comparison of the short time decay for a double well
condensate which has 200 atoms equally distributed to two wells.
Solid line corresponds to a coherent state, dashed line corresponds
to a squeezed state with $\zeta=0.9$ and dotted lines to squeezed
states with $\zeta=0.5i$ and $\zeta=-0.5$.}
\end{figure}

When the results for single and double well systems are compared, it
is seen that although decay times are both in the same order of
magnitude, a double well condensate with equal number of atoms in
each well as a single well condensate suffers faster decay. This
agrees with the analytical results obtained in \cite{toy1} using the
same ansatz for single and double well ground states.

\section{Thin spectrum formalism}
\subsection{Introduction}
By thin spectrum, we refer to a group of states whose energy
spacings are so low that they are not controllable by any experiment
and whose effect on the free energy becomes zero in the
thermodynamic limit. That the existence of a thin spectrum leads to
decoherence of excitations at finite temperature is proved in
\cite{wezel05,wezel06}. In \cite{wezel05} it is shown that
excitations on a Lieb-Mattis system suffer decoherence with a rate
proportinal to $k_B T / N \hbar$ where $N$ is the system size. In
\cite{wezel05} the same authors claim that this time scale, being
independent of the details of the system, applies to other physical
systems too, and in \cite{wezel07} they prove that Hubbard model
superconductors suffers the same decay. In this section, we apply
the thin spectrum formalism to atomic BECs and show that they suffer
collapse in the same time scale.

\subsection{Quasiparticle lifetime in Bose-Einstein condensates}
We now go back to the Hamiltonian (\ref{equ_ham1})
\begin{eqnarray}
\mathcal{H}=\sum_k E_k a_k^\dagger a_k + \frac{\tilde
u}{2}\sum_{k,p,q}a_{p+q}^\dagger a_{k-q}^\dagger a_k a_p.
\end{eqnarray}
Omitting the $3^{rd}$ and $4^{th}$ order terms in the non-condensed
modes (${k\neq 0}$) we get
\begin{equation}
\mathcal{H} = \mathcal{H}_z + \mathcal{H}_e,
\end{equation}
\begin{equation}
\mathcal{H}_z = \frac{\tilde u}{2} (\hat{n}_0^2 -\hat{n}_0),
\end{equation}
\begin{equation}
\mathcal{H}_e = \sum_{k\neq0}\left[\left(E_k +2\tilde{u}
            \hat{n}_0\right)\hat{n}_k +
            \frac{\tilde u}{2}\left(a_k^\dagger a_{-k}^\dagger a_0 a_0 +
            h.c.\right)\right].
\end{equation}
In order to study the excitations and zero mode separately, we need
$[\mathcal{H}_z,\mathcal{H}_e]=0$ and for this we neglect the
quantum nature of $a_0$ in $\mathcal{H}_e$ by replacing
$\hat{n}_0/V$ appearing in $\mathcal{H}_e$ by $\rho_0 = N_0 /V$.
Here $N_0$ denotes the number of atoms in the zero mode, and so
$\rho_0$ is the corresponding density. After substituting the
chemical potential that gives the correct number of atoms,
$\mu_0=u_0 \rho_0 -u_0\rho_0/2N_0$, we get
\begin{equation}
\mathcal{H}=\frac{u_0 \rho_0}{2 N_0} \hat{n}_0^2 -\rho_0 u_0
\hat{n}_0 + \mathcal{H}_e .
\end{equation}
\begin{eqnarray}
\mathcal{H}_e=\sum_{k \neq 0}\left[\epsilon_k \hat{n}_k + \frac{u_0
\rho_0}{2}(a_k^\dagger a_{-k}^\dagger + h.c.)\right]. \label{Hbg}
\end{eqnarray}
Here, $u_0$ is the interaction strength not scaled with V, that is
$u_0= V \tilde{u}$, and $\epsilon_k$ is defined as $\epsilon_k =
E_k+2u_0\rho_0$. The excitation Hamiltonian can be diagonalized to
give
\begin{eqnarray}
\mathcal{H}_e=\sum_{k \neq 0} \omega_k b_k^\dagger b_k + const.,
\end{eqnarray}
with $\omega_k=[\epsilon_k^2 - u_0^2 \rho_0^2]^{1/2}$ \cite{greiner}
and $b_k=S a_k S^{-1}$. $S$ is the multi-mode squeeze operator
\cite{haque06}.

For simplicity, we consider a system with quasiparticle excitations
in only one mode. We denote such a system with $n$ atoms in the
condensate and $m$ quasiparticles with $\omega$ by $|n,m\rangle$.
Then,
\begin{eqnarray}
\hat{n}_0|n,m\rangle &=& n |n,m\rangle,\\
\hat{n}_{k'}|n,m\rangle &=& m\,\delta_{k,k'} |n,m\rangle ,
\end{eqnarray}
\begin{equation}
\mathcal{H}|n,m\rangle = E_m^{(n)} |n,m\rangle,
\end{equation}
We consider single-particle regime such that quasiparticle and
particle occupation numbers become the same. Due to the number
conservation, sum of condensate atoms and the quasiparticles
should remain the same. To excite $m$ quasiparticles, we have to
decrease condensate atom number by $m$. The energy of the
corresponding state becomes
\begin{equation}
E_m^{(n)} = \left[\frac{u_0 \rho_0n^2}{2(N_0 - m)} - u_0 \rho_0 n +
m\omega\right].
\end{equation}

Following \cite{wezel06}, we assume that in the beginning the system
has no quasiparticle excitations at all, and therefore has a
Boltzmann weighted distribution over the states $|n,0\rangle$, i.e.,
\begin{eqnarray}
\rho(t=0)\propto\sum_n {\rm e}^{- \beta E_0^{(n)}}
|n,0\rangle\langle n,0|. \label{equ42}
\end{eqnarray}
This state has no phase, and therefore is not the perfect starting
point for a BEC. However, if we are interested only in the collapse
of the excitations and if this take place on a time scale smaller
than the time of phase diffusion of the zero mode, then we can get
an estimate. For the time being we will study excitations on this
thermal state, and in the following subsection we will generalize
our ideas to thermal coherent states.

Before proceeding further, we replace the sum in eq. (\ref{equ42})
by an integral. Since the value of $E_n$ will be extremely small for
$n<0$, it is also legitimate to expand this integral to include the
negative values of n too.

Now we bring the system to a superposition of the
zero-quasiparticle state and the one quasiparticle state, that is
we bring each $|n,0\rangle$ to
$(|n,0\rangle+|n,1\rangle)/\sqrt{2}$. Such a state can be
interpreted as a particular qubit \cite{wezel06}. For thin
spectrum to affect the system, it is essential that we bring the
system to a superposition rather than simply exciting a
quasiparticle. Now, the off diagonal element of the resulting
state's density matrix will evolve according to
\begin{eqnarray}
\rho_{\rm od}(t>0) &\propto& \int_{-\infty}^{\infty}  {\rm
e}^{-\beta E_0^{(n)}}
{\rm e}^{-\frac{i}{\hbar}(E_1^{(n)} - E_0^{(n)}) t} dn \nonumber\\
&\propto& \int_{-\infty}^{\infty} {\rm e}^{(-\beta u_0
                \rho_0/2 N_0 + i t u_0 \rho_0/2\hbar N_0^2)n^2 +
                \beta \rho_0 u_0 n} dn \nonumber\\
                &\propto& \sqrt{\pi} \frac{\exp\left(\frac{\beta^2
                \rho_0^2 u_0^2}{2\beta u_0 \rho_0 / N_0 - 2 i t u_0
                \rho_0 / \hbar N_0^2}\right)}{\sqrt{\beta u_0 \rho_0 / 2N_0 - i t u_0 \rho_0 / 2 \hbar
                N_0^2}},
\end{eqnarray}
which gives
\begin{eqnarray}
|\rho_{\rm od}(t)|^2 \propto \frac{\exp\left(\frac{\beta^3 N_0^3 u_0
\rho_0}{\beta^2 N_0^2 + t^2/\hbar^2} \right)}{\sqrt{\beta^2 + t^2
/\hbar^2 N_0^2}}, \label{decay_4}
\end{eqnarray}
after omitting terms with only a phase factor. Although the
denominator and the numerator have quite different forms, we find
that both decay in a time proportional to $t_{c}\sim \hbar N_0 / k_B
T$. This is the same result that Wezel \textit{et. al.} have found
for a crystal \cite{wezel06}.

For a BEC, we can let $N_0\sim 10^6 - 10^8$ and $T\sim 10^{-8} -
10^{-7}$ K. This gives $t_{c} \sim 10^2 - 10^5$ seconds, which is
much larger than even the ground state life times. Unlike the room
temperature mesoscopic system discussed in \cite{wezel06}, BECs are
extremely cold systems therefore one single excitation has such a
long life time. However, this does not make the calculation
unuseful. This life time is given only for a single quasi-particle
excitation. In general it might be useful to have more than one
quasiparticles excited at a time. If, for example, the condensate
will be used as the building block of a quantum computer, having an
excitation consisting of $\sim N$ quasiparticles will make
observation of the qubit (superpositions of states with $m=0$ and
$m$ quasiparticles), easier. In order to find a decay time for
$m>1$, the only approximation required is $1/N(N-m) \simeq 1/N^2$
and the timescale will be inversely proportional to m. For $m \sim
N_0$, so long as it is not the case that $1-m/N\ll 1$, we have
$t_{c} \sim 10^{-4} - 10^{-3}$ seconds. This time scale is much
smaller than both the observed and expected ground state life times,
therefore is of interest.

There are different studies \cite{gora,liu} concerning the life
times of quasiparticle excitations, such as using perturbation
theory, etc. Namely, \cite{gora} has found a linear temperature
dependence for high energy quasiparticles. Our calculation will make
a quantitative contribution to this decay rate. However, low energy
excitations are shown to have more complex temperature dependencies
\cite{liu}. Our calculations do not make any predictions for that
regime, since we have assumed $E_k\gg u_0\rho_0$. Also, our theory
predicts a certain dependence of the life time on the number of
quasiparticles excited (which is almost linear for small $m$), and
this might be used to differentiate it from other theories.

As discussed before, for a group of states to be a thin spectrum,
their effect on the free energy must vanish. For this purpose, we
write the partition function as
\begin{equation}
Z=Z_{thin} \cdot Z_{observable}
\end{equation}
with $Z_{thin}=\sum_n {\rm e}^{-\beta u_0 \rho_0 (n^2/2N_0 -n)}$.
Again replacing the sum by an integral we find that the leading term
in the free energy per particle $\ln(Z_{thin})/N_0 \sim
\ln(N_0)/N_0$. This means that the mode we consider has no effect on
free energy and satisfies all the criteria to constitute a thin
spectrum.

The next step might be to generalize the calculations for coherent
of squeezed coherent zero mode occupations. However, following this
path doesn't give any finite life time, since the excitation decay
due to thin spectrum requires finite temperature, but coherent
states have no temperature characteristic. But it is natural to do
the same calculations for thermal coherent zero mode occupation.
Such a calculation is presented in \cite{myself}. Decay rate does
not have a linear temperature dependence in this case, therefore it
might be possible to differentiate between thermal and thermal
coherent occupations experimentally.

\subsection{Effect of Thin Spectra on a Double Well Condensate}
In \cite{toy1}, it is shown that the phase related part of the
Hamiltonian of a double well condensate can be reduced to the form
\begin{equation}
\mathcal{H}=\alpha_+ P_+^2 + \alpha_- P_-^2 + \lambda \gamma_- Q_-^2
.
\end{equation}
Here, $\alpha_+$, $\alpha_-$ and $\gamma_-$ are parameters depending
on system details, $P_+$ and $P_-$ are the momenta corresponding to
the total and relative phases of the condensates, and $Q_-$ is the
coordinate corresponding to the relative phase. $\lambda$, again, is
a number parameterizing the tunneling between different wells. (This
Hamiltonian, originally derived for a condensate consisting of two
different types of atoms in a single potential well, is applicable
to a double well system when the parameter corresponding to
collisions between different types of atoms is taken to be zero.)
When coupling $\lambda$ between the wells is taken to be zero, this
Hamiltonian reduces to one of two free particles:
\begin{equation}
\mathcal{H}=\alpha_+ P_+^2 + \alpha_- P_-^2 .
\end{equation}
It is seen that the system has two modes corresponding to motions
without restoring forces. The reason is that now that there are two
wells, there is an extra symmetry that is spontaneously broken.
Therefore, there are two different thin spectra.

Cumulative effect of multiple broken symmetries (and hence multiple
thin spectra) on excitation lifetime is studied in \cite{myself}. If
lifetime corresponding to individual thin spectra are $t_1$ and
$t_2$, then the resultant collapse time is the harmonic sum of
individual life times:
\begin{equation}
t_r^{-1}=t_1^{-1}+t_2^{-1}.
\end{equation}
Using $\alpha$'s corresponding to the system under consideration, it
is thus possible to find the resultant life time, which is supposed
to be in the same order of magnitude with the smaller life time.

\section{Conclusions}
Generalizing the Toy model calculations \cite{toy1,toy2}, we
discussed the phase decoherence of coherent, squeezed coherent and
thermal coherent states. For visual clarity, time dependence of
various Q functions is shown. A generalization of the toy model to
double well systems is also discussed and time evolution of the
order parameter is studied for coherent and squeezed coherent
states. This step, being important not only for double well BECs,
might bear important results for any Josephson-coupled system.

The effect of thin spectrum \cite{wezel06,wezel05} on quasiparticle
excitations in BECs is briefly reviewed. It is shown that the
presence of the so called thin spectrum states, which have vanishing
level spacing, also has no effect on free energy per particle in the
thermodynamic limit. Qualitative dependence of life time on the
number of excitations is given. Finally, as a simple example of a
system with more than one spontaneously broken symmetry
\cite{myself}, a calculation of excitation life time in a double
well system is outlined.

\section{Acknowledgements}

T.B. is supported by T\"UB\.ITAK. O.E.M. acknowledges the support
from a T\"UBA/GEB\.{I}P grant. T.B. acknowledges fruitful
discussions with Jasper van Wezel and Patrick Navez.

\end{document}